\documentclass{PoS}
\usepackage{amsmath}
\usepackage{graphicx}

\title{Can Pulsars in the inner parsecs from Galactic Centre probe the existence of Dark Matter?}
\ShortTitle{Pulsars in GC as DM probes}

\author{Antonino Del Popolo\\
     Dipartimento di Fisica e Astronomia, University Of Catania, Viale Andrea Doria 6, I-95125 Catania, Italy.\\
     INFN Sezione di Catania, Via S. Sofia 64, I-95123 Catania, Italy.\\
     Institute of Astronomy, Russian Academy of Sciences, Pyatnitskaya str. 48, 119017 Moscow, Russia.\\
      E-mail: \email{adelpopolo@oact.inaf.it}}
\author{\speaker{Maksym Deliyergiyev} \\
     Institute of Physics, Jan Kochanowski University, PL-25406 Kielce, Poland.\\
      E-mail: \email{maksym.deliyergiyev@ujk.edu.pl}}
\author{Morgan Le Delliou\\
     Institute of Theoretical Physics, School of Physical Science and Technology, Lanzhou University, No.222, South Tianshui Road, Lanzhou, Gansu 730000, China.\\
     Instituto de Astrof\'isica e Ci\^encias do Espa\c co, Universidade de Lisboa, Faculdade de Ci\^encias,Ed. C8, Campo Grande, 1769-016 Lisboa, Portugal.\\
      E-mail: \email{delliou@lzu.edu.cn,delliou@ift.unesp.br}}
\author{Laura Tolos\\
Institut f\"{u}r Theoretische Physik, Goethe Universit\"{a}t Frankfurt, Max-von-Laue-Stra\ss{}e 1, 60438 Frankfurt, Germany.\\
Frankfurt Institute for Advanced Studies, Goethe Universit\"{a}t Frankfurt, Ruth-Moufang-Str.1, 60438 Frankfurt am Main, Germany.\\ 
Institute of Space Sciences (ICE, CSIC), Campus UAB, Carrer de Can Magrans, 08193, Barcelona, Spain.\\
Institut  d'Estudis Espacials de Catalunya (IEEC), 08034 Barcelona, Spain.\\
      E-mail: \email{tolos@th.physik.uni-frankfurt.de}}
\author{Fiorella Burgio\\
INFN Sezione di Catania, Via S. Sofia 64, I-95123 Catania, Italy.\\
      E-mail: \email{fiorella.burgio@ct.infn.it}}

\abstract{
We discuss the formation of dark compact objects in a dark matter environment in view of the possible mass dependence of pulsars on the distribution of dark matter in the Galaxy. Our results indicate that the pulsar masses should decrease going towards the center of the Milky Way due to dark matter capture, thus becoming a probe for the existence and nature of dark matter. 
We thus propose that the evolution of the pulsar mass in a dark matter rich environment can be used to put constraints, when combined with future experiments, on the characteristics of our Galaxy halo dark matter profile, on the dark matter particle mass and on the dark matter self-interaction strength.
}

\FullConference{The European Physical Society Conference on High Energy Physics (EPS-HEP), 10-17 July 2019\\
		Ghent, Belgium}

\begin{document}

\section{Introduction}

Dark matter (DM) is a key ingredient for models
that try to explain 
cosmological structure formation without modifying gravity. Although the gravitational effects of DM are well documented \cite{Betoule:2014frx,Ade:2013zuv},
direct detection of particles for this dominant matter component continues to elude proofs: in accelerators or in nuclear recoil experiments 
\cite{
Felcini:2018osp,
Klasen:2015uma},
indirect WIMP annihilation searches \cite{Conrad:2014tla}, in DM stars \cite{Dai:2009ik,Kouvaris:2015rea} or in some other indirect quests as illustrated in Refs.~\cite{
Bertolami:2012yp,
Abdalla:2007rd,
Delliou:2014awa}.

In this context, different testing avenues of possible DM effects are welcome, such as in pulsars, i.e. rotating neutron stars (NSs), which provide the advantage of extreme densities and can accrete DM, thus straining  the saturated neutron gas. The amount of DM acquired by a NS follows the Tolman-Oppenheimer-Volkoff (TOV) equation 
, as in e.g. \cite{Tolos2015}. 
Moreover, the effect of DM on NSs can directly lead to bounds for the masses of the different DM candidates \cite{Goldman:1989nd,Kouvaris:2011fi}.

Self-annihilating DM can also produce characteristic effects on NS
\cite{Kouvaris2008,
Bertone2008,
Kouvaris2010,
McCullough:2010ai,
deLavallaz2010,
PerezGarcia:2011hh,
PerezGarcia:2010ap}.
In particular, WIMPs annihilation in DM cores should produce temperature and luminosity changes, through heat, of old stars \cite{Kouvaris2008,Bertone2008,Kouvaris2010,deLavallaz2010}. 
However, those changes  are 
difficult to 
detect \cite{Kouvaris2008,Sandin:2008db}. 

In this short report, based on the results obtained in our previous work \cite{Deliyergiyev:2019vti} and on our recent paper \cite{DelPopolo:2019nng}, we propose a testable galactic probe for DM existence in the form of evolution of the pulsar mass towards the galactic centre (GC). According to the discussion above, NSs in increasingly DM rich environments should accrete more DM and thus display a characteristic mass decrease, the closer they are to the galactic centre. 
We use NSs because first of all the very large baryon density inside NSs makes the interaction between baryons and DM following DM capture most likely; second, the NSs strong gravitational force makes DM particles escape very unlikely, after they interact and loose energy. This mass evolution is easier to test than other probes such as NS temperature time evolution with DM accretion, as discussed previously.

\section{Mass Change of Neutron Stars}
\label{sec:Implementation}

\begin{figure}[t]
  \begin{center}
    \includegraphics[width=0.395\textwidth]{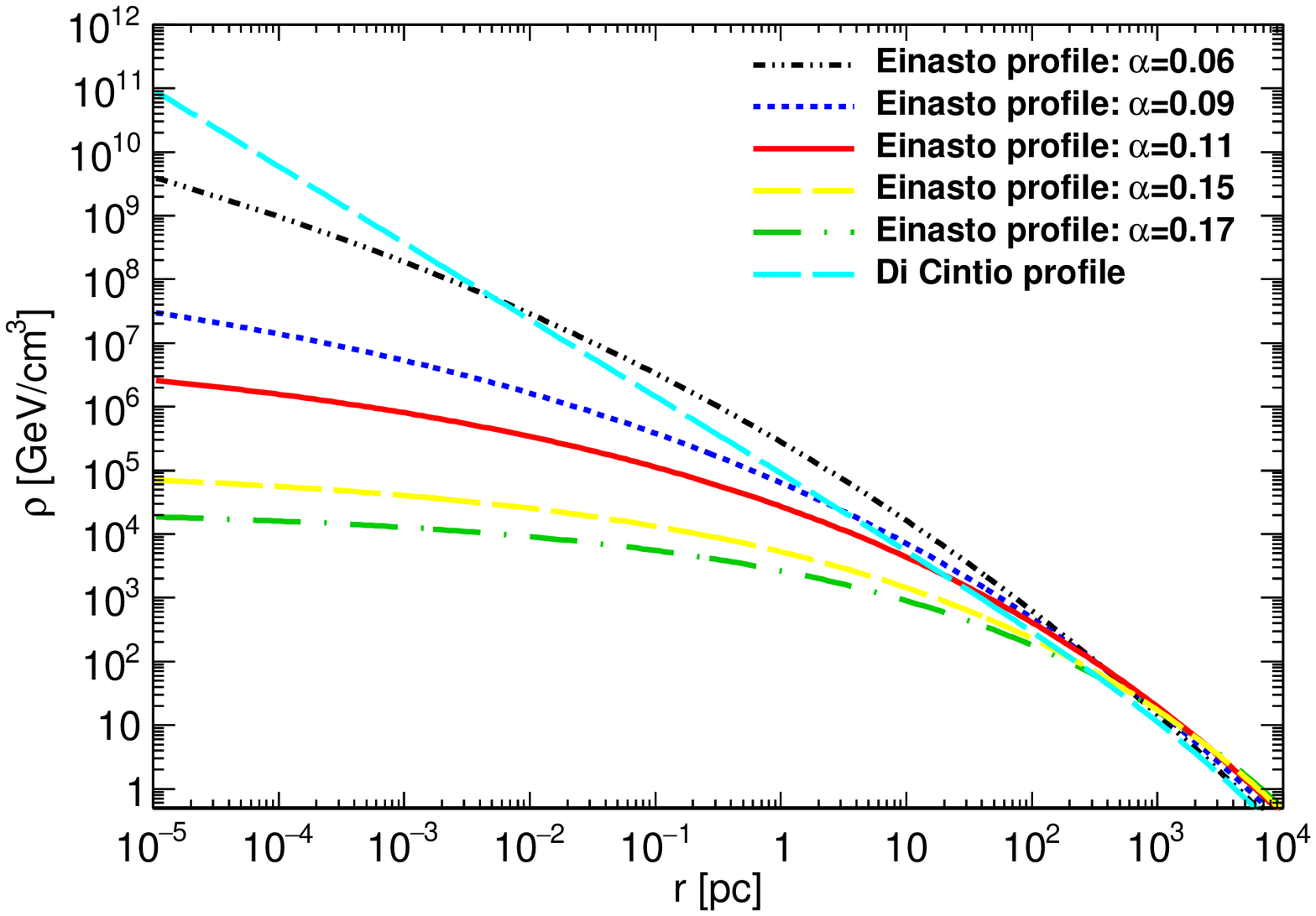}
    \includegraphics[width=0.395\textwidth]{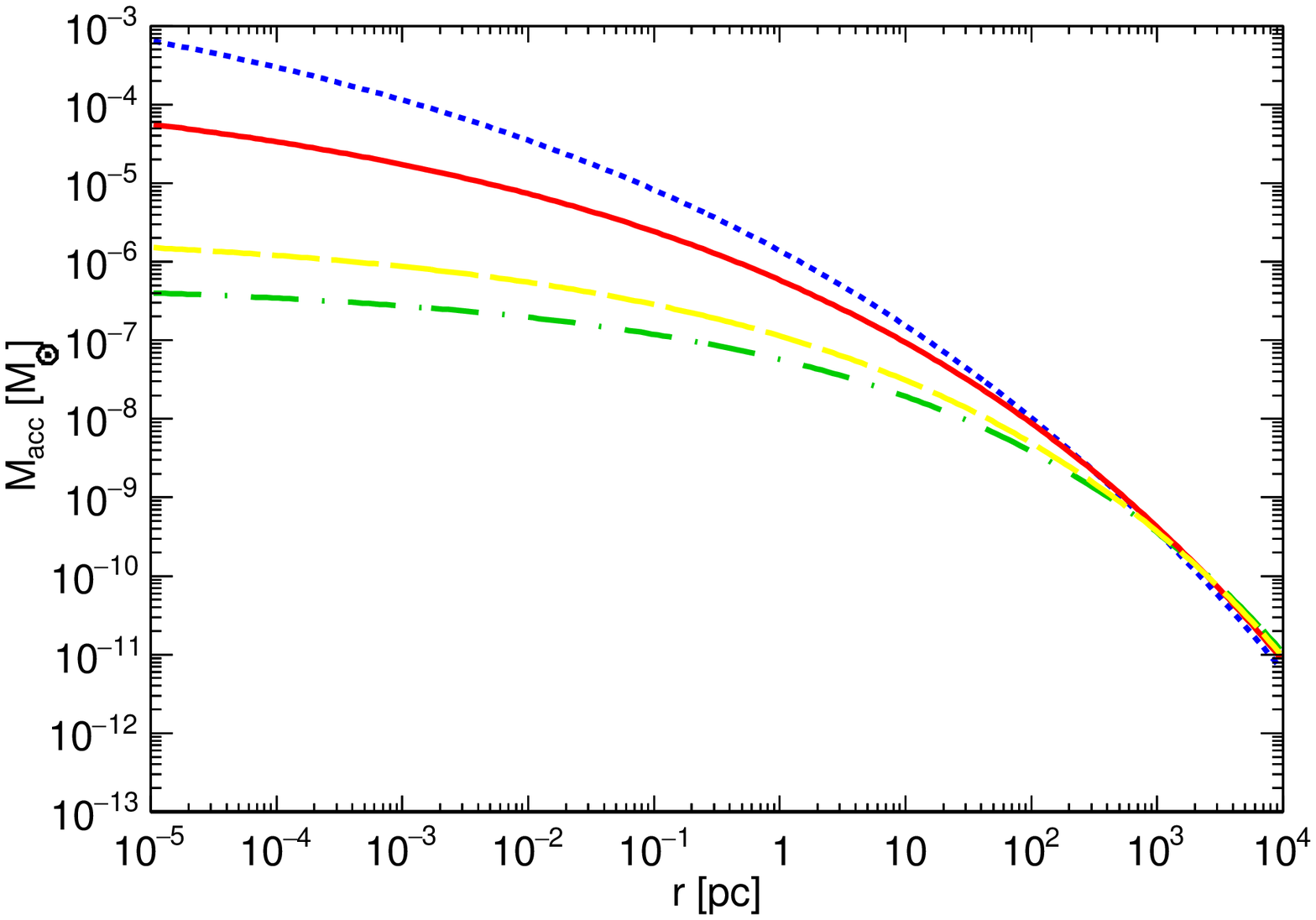}
 \caption{
Figures taken from Ref.~\cite{DelPopolo:2019nng}. 
(Left panel) Einasto profile for $\alpha=0.06,0.09,0.11,0.15,0.17$ (black dot dot dashed, blue dotted, red solid, yellow long dashed, and green dot dashed lines, respectively). The cyan dashed line is the Di Cintio profile \cite{DiCintio:2014xia}  (labelled DC14 in the text). (Right panel) The accreted mass according to 
Kouvaris formula (Eq.~(\ref{eq:Kouv2013})). In this plot we do not consider the accreted mass corresponding to the Di Cinto profile, nor to 
the Einasto profile with $\alpha=0.06$, in order to be conservative in our calculations. 
}
\label{fig:einasto}
  \end{center}
\end{figure}

As shown by Ref.~\cite{Kouvaris2013}, in the framework of a spherically symmetric accretion scenario for a typical NS of mass $1.4 M_\odot$ and $R=10$ km, 
the total accreted mass is given by Ref.~\cite{DelPopolo:2019nng} 
\begin{equation} 
M_{\rm acc}= 1.3 \times 10^{43} \left( \frac{\rho_{\rm dm}}{\rm 0.3 \, GeV/cm^3} \right) \left(\frac{\rm t}
{\rm Gyr} \right) f \,\,\, \rm GeV,
\label{eq:Kouv2013}
\end{equation}
which is an underestimation of a factor $\simeq 10$, since the accretion during the NS progenitor phase, of the same order as in the NS phase \cite{Kouvaris2010} (factor of 2), and the accretion coming from DM self-interaction \cite{Guver2014} are not taken into account. 
Moreover, $M_{acc}$ of Eq.~(\ref{eq:Kouv2013}) should be corrected by factor 2.05 for 2 $M_{\odot}$

We obtain DM accretion $\simeq 10^{-11} {\rm M}_{\odot}$ for a typical NS in the solar neighborhood, 
which is in agreement with the capture rates of Refs~\cite{Kouvaris2013, Guver2014, Zheng2016, Zhong2012} and the results from  \cite{Kouvaris2011}, but below the estimates from the DM accumulated using TOV. 
A better agreement between the accreted DM mass and the accumulated DM  mass coming from TOV 
is obtained for NSs located in Superdense DM clumps, Ultra Compact mini-haloes \cite{Berezinsky2013}, and close to the GC.

The another approach starts, as in usual structure formation, from linear perturbation, followed by non-linear collapse of DM, and then continued by baryons collapse in the DM potential wells previously formed \cite{Chang2018}.

\begin{figure}[t]
  \begin{center}
    \includegraphics[width=0.95\textwidth]{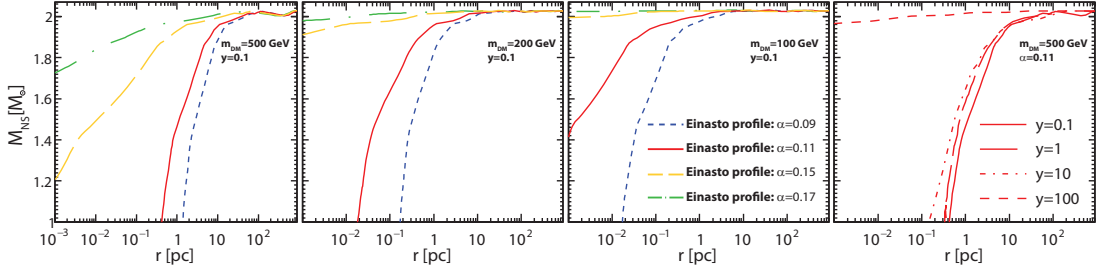}
    \caption{
Figures taken from Ref.~\cite{DelPopolo:2019nng}. 
The changes of the NS mass from DM accumulation as a function of the NS distance to GC, for a particle mass $m_{DM}=500$ (leftmost), 200 (center left) and 100 (center right) GeV, with $y=0.1$ and  $\alpha=0.09; 0.11; 0.15; 0.17$ (blue dotted, red solid, yellow long dashed, and green dot dashed). The rightmost panel shows the role of the interaction strength for particle mass $m_{DM}=500$ GeV, $\alpha=0.11$, and $y=0.1, 1, 10, 100$ (solid, long dashed, dash dotted, and long dashed dotted lines).
    }
   \label{fig:accreteMass}
  \end{center}
\end{figure}

Once the accreted DM mass, $M_{acc}$, as a function of the distance to the GC is determined from the right panel of Fig.~\ref{fig:einasto}, we can determine the corresponding mass change of the NS towards the galactic center. 
In order to do so, we use Fig.10 of our previous work \cite{Deliyergiyev:2019vti}, where the maximum mass of NSs was obtained as a function of the DM mass inside the NS, $M_{DM}$, for the DM weakly interacting case, $y=0.1$\footnote{ The interaction strength is expressed in terms of the ratio of the DM fermion mass $m_f$, and scale of interaction $m_I$, $y=m_f/m_I$. This can be converted to usual units: one can estimate the cross section of DM self-interaction, taking the mass of the DM particle $m_f$ in units of GeV, as 
\begin{align}
\sigma =& \frac{1}{4\pi}\frac{m_f^2}{m_I^4}
= \frac{y^4}{m_f^2} 3.2\times 10^{-27} \textrm{cm}^2  &
\rightarrow \sigma/m_f =& \frac{y^4}{m_f^3} 1.8\times 10^{-3} \frac{\textrm{cm}^2}{g} 
\label{eq:yCrossSection}
\end{align}
}. 
Moreover, we obtained similar plots for larger values of the interaction parameter $y$, $y=1,10,100$. Since $M_{DM}$ must be equal to the accreted mass, $M_{acc}$, we find a relation between the total mass of the NS, $M_{NS}$ ($M_T$ in the notation of \cite{Deliyergiyev:2019vti}) and the distance from the GC. The result is plotted in Fig.~\ref{fig:accreteMass}. 
Its leftmost panel displays the change in mass of a NS moving towards the Milky Way (MW) halo center (colour and line coding as in Fig.~\ref{fig:einasto}), for a particle mass, $m_{DM}=500$ GeV, and $y=0.1$. 
Our reference model ($\alpha=0.11$, red solid line) shows a NS mass change from 2 to 1 $M_{\odot}$ at 0.4 pc, while in the case $\alpha = 0.9$ (blue dotted line) the same change is observed at 1.35 pc. The other two cases show a slower mass change. 
The yellow long dashed line ($\alpha = 0.15$), and the green dot dashed line ($\alpha = 0.17$) show that the mass reduces from 2 to 1.2 $M_{\odot}$, and from 2 to 1.7 $M_{\odot}$, at $10^{3}$ pc, respectively. 
The next two panels show how the mass changes with decreasing the DM particle mass ($m_{DM}=200$ GeV, centre left, 100 GeV, centre right). Finally the rightmost plot shows the effect of the interaction strength for our reference profile ($\alpha=0.11$), and from right to left, in the cases $y=0.1;1;10;10;100$. Strong interaction ($y=100$) produces very small mass changes, while the weaker the interaction, the larger the change is.

All the above results are obtained with conservative assumptions, i.e.
(a) not taking into account
(i) NS progenitor accretion, expected of order of the NS phase \cite{Kouvaris2010},
(ii) DM self-interaction accretion \cite{Guver2014},
(b) using a density profile from DM-only simulations, shallower than in recent hydrodynamic simulations \cite{DiCintio:2014xia}.

As the right panel of Fig.~\ref{fig:einasto} shows, we excluded from the analysis the $\alpha = 0.06$ Einasto profile, and even the DC14 profile \cite{DiCintio:2014xia}, a realistic profile calculated with hydro-dynamical simulations, to remain conservative. Although we made this choice, the orbital dynamics of PSR B1257+12 \cite{Iorio:2010hb}, together with the accretion predictions \cite{Bertone:2004pz} allows much larger DM accumulation in NS than in our present work: up to 10\% of a NS mass, in agreement with DC14 \cite{DiCintio:2014xia}. Moreover, complex astrophysical phenomena are occurring on sub-parsec scale near the GC, such as DM particles gravitational scattering by stars and capture in the supermassive BH, together with highly enhanced central density from the supermassive BH formation \cite{Gondolo1999}. Thus, more accurate density profiles can be introduced \cite{Bertone:2004pz}.

\section{Conclusion}

We have shown that the NS mass should reflect the changes of the DM environment in the MW(Fig.~\ref{fig:accreteMass}). This is done taking into account the DM accretion of NSs [see \cite{Deliyergiyev:2019vti}, for details], as it changes because of the increase of DM content \cite{Kouvaris2013} when we move towards the GC \cite{Bernal:2011pz}. This allows us to propose that the evolution of the pulsar
masses towards the GC of the MW can be a probe of the existence of DM. In fact, the decrease of the NS mass for NSs located closer and closer to the GC would put
constraints on the characteristics of the Galaxy halo dark matter profile, on the dark matter particle mass, and on the self-interaction strength. Such changes are expected to be observed in the near future in telescopes such as ngVLA \cite{murphy2018science}, SKA \cite{Konar:2016lgc}, Athena \cite{barcons2012athena, EESA2014athena}, NICER \cite{10.1117/12.926396} or eXTP \cite{Watts2018}.

\section{Acknowledgements}
M.D. work was supported by the Polish National Science Centre (NCN) grant 2016/23/B/ ST2 / 00692. MLeD acknowledges the financial support by Lanzhou University starting fund and the Fundamental Research Funds for the Central Universities (Grant No.lzujbky-2019-25). L.T. acknowledges support from the FPA2016-81114-P Grant from Ministerio de Ciencia, Innovacion y Universidades, Heisenberg Programme of the Deutsche Forschungsgemeinschaft under the Project Nr. 383452331 and PHAROS COST Action CA16214.

\end{document}